\documentclass{acta}

\usepackage{amsmath}
\usepackage{amssymb}
\usepackage{rotating}
\usepackage[hidelinks]{hyperref} %%PM

\usepackage{xcolor}

\newcommand\tE{t_{\rm E}}
\newcommand\thetaE{\theta_{\rm E}}

\newcommand\jkas{J. Korean Astron. Soc.}

\newcommand\fs{\mbox{$.\!\!^{\mathrm s}$}}% 
\newcommand\arcdeg{\mbox{$^\circ$}}% 
\newcommand\arcmin{\mbox{$^\prime$}}% 
\newcommand\farcs{\mbox{$.\!\!^{\prime\prime}$}}% 

\begin{document}

\begin{Titlepage}

\Title{\textit{TESS} Free-floating Planet Candidate Is Likely a Stellar Flare}

%\correspondingauthor{Przemek Mr\'oz}
%\email{pmroz@astro.caltech.edu}

\Author{P~r~z~e~m~e~k~~~M~r~{\'o}~z}{Astronomical Observatory, University of Warsaw, Al. Ujazdowskie 4, 00-478 Warszawa, Poland\\ e-mail: pmroz@astrouw.edu.pl}

\Received{MM DD, YYYY}

\end{Titlepage}

\Abstract{The discovery of a terrestrial-mass free-floating planet candidate in the light curve of the star TIC~107150013 observed by the \textit{Transiting Exoplanet Survey Satellite} (\textit{TESS}) has recently been announced. A short-duration ($\approx 0.5$ day), low-amplitude ($\approx 0.06$ mag) brightening in the \textit{TESS} light curve was interpreted as a short-timescale gravitational microlensing event. However, the purported event occurred far from the Galactic center and the Galactic plane ($l\approx 239^{\circ}$, $b\approx -5^{\circ}$), on a relatively nearby ($\approx 3.2$ kpc) star, making the microlensing interpretation unlikely. Here, we report the archival photometric observations of TIC~107150013 collected by the Optical Gravitational Lensing Experiment (OGLE) from 2018 through 2020. The archival OGLE light curve reveals periodic variability indicative of starspots on the surface of the star. The presence of starspots indicates magnetic activity of the star, which may also manifest as stellar flares. We interpret the brightening of TIC~107150013 seen in the \textit{TESS} data as the stellar flare. We present similar flaring stars detected in the archival OGLE data, mimicking short-timescale, low-amplitude microlensing events. Such stars may be a source of non-negligible false positive detections in the planned space-based microlensing surveys.
}
{Gravitational microlensing (672), Free-floating planets (549), Stellar activity (1580), Starspots (1572), Stellar flares (1603), Flare stars (540)}

\section{Introduction} \label{sec:intro}

High-cadence photometric observations by ground-based gravitational micro\-len\-sing surveys have uncovered a population of short-timescale microlensing events \citep{mroz2017,gould2022,sumi2023} that are interpreted as due to free-floating (or wide-orbit) planets. These studies indicate that free-floating planets are common (about $7^{+7}_{-5}$ objects more massive than $1\,M_{\oplus}$ per star) and that their mass function approximately scales as $dN/d\log M \propto M^{-1}$.
We refer the readers to reviews by \citet{zhu_dong_2021} and \citet{mroz_poleski_2024} for an in-depth discussion of these results.

Microlensing events by terrestrial-mass objects are expected to have short Einstein timescales (usually a few hours) and small amplitudes due to finite-source effects, rendering them difficult to detect in ground-based data. Therefore, even though the population of free-floating planets seems to be large, only a handful of candidates have been detected by ground-based microlensing surveys in the past years.

The sensitivity to short-timescale microlensing events can be boosted with nearly continuous, high-cadence observations conducted from space. For example, the planned \textit{Nancy Grace Roman Space Telescope} is expected to detect hundreds of short-timescale microlensing events due to free-floating planets as part of its Galactic Bulge Time Domain Survey \citep{johnson2020,sumi2023}. Similarly, the \textit{Earth 2.0} (\textit{ET}) mission is estimated to find about 600 free-floating planets \citep{ge2022}. These missions were designed to observe dense stellar fields of the Galactic bulge at high cadence.

\citet{kunimoto2024} reported the results of searching for short-duration microlensing events in the photometric data from one of the sectors (Sector 61) observed by the \textit{Transiting Exoplanet Survey Satellite} (\textit{TESS}). \textit{TESS} observed a field of view $24^{\circ} \times 96^{\circ}$ wide for 27.4 days at a 200-second cadence. The analyzed data set contained 1.3 million stars as faint as $T=15$ mag. \citet{kunimoto2024} used the modified version of the BLS period search algorithm to select candidate microlensing events. They vetted all selected objects by checking the quality of the microlensing model fit, removing edge artifacts, asteroids, and repeating or asymmetric light curves. 

Only one object, TIC~107150013 ($\mathrm{RA}=07^{\rm h}22^{\rm m}30\fs 81$, $\mathrm{Decl.}={-25}\arcdeg{29}\arcmin{15}\farcs{7}$, J2000), met all selection criteria and was considered to be a compelling microlensing event candidate. The brightening observed by \textit{TESS} lasted about $\approx 0.5$ day and had a low amplitude ($\approx 0.06$ mag). The best-fit microlensing model fitted to the light curve yielded the Einstein timescale $\tE = 0.074 \pm 0.002$ days, and the angular Einstein radius $\thetaE = 4.1^{+0.4}_{-0.5}\,\mu\mathrm{as}$, which are similar to those of other known free-floating planet candidates.

All known microlensing events due to free-floating planets have occurred toward the dense regions of the Galactic bulge, where the probability of finding microlensing events is highest. On the contrary, the star TIC~107150013 is located at the Galactic coordinates $l=239.1^{\circ}$, $b=-5.0^{\circ}$, at a distance of $3.19 \pm 0.15$ kpc \citep{kunimoto2024}. The probability of observing a microlensing event so far away from the Galactic center and the Galactic plane, on such a nearby star, is extremely low. \citet{mroz2020c} found that the microlensing event rate in the Galactic plane decreases exponentially with the increasing angular separation from the Galactic center $\Gamma \propto \exp(-|l|/l_0)$, where $l_0 = 31.5^{+4.2}_{-3.7}$ deg. At $l = -120.9^{\circ}$, the average event rate is approximately $0.6\times 10^{-7}\,\mathrm{yr}^{-1}\,\mathrm{star}^{-1}$, and is $\approx 300$ times lower than that toward the Galactic bulge. The estimated number of microlensing events in one \textit{TESS} sector is approximately $0.6 \times 10^{-7} \times 1.3 \times 10^6 \times 27.4 / 365.25 \approx 0.006$, assuming 100\% detection efficiency. The expected number of microlensing events is even lower because \citet{mroz2020c} calculated the event rate averaged over the latitude range $-7^{\circ} < b < 7^{\circ}$ and $\Gamma$ is known to be decreasing with the increasing distance from the Galactic plane, the analyzed \textit{TESS} star is relatively nearby, and events due to free-floating planets are expected to be far less frequent than those by stars. {\color{black} Indeed, a detailed calculation by \citet{yang2024} found that only 0.0018 microlensing events due to free-floating planets are expected to be detected in \textit{TESS} Sector 61.} That makes the interpretation that TIC~107150013 showed a short-duration microlensing event very unlikely. 

\begin{figure}[htb]
\centering
\includegraphics[width=.8\textwidth]{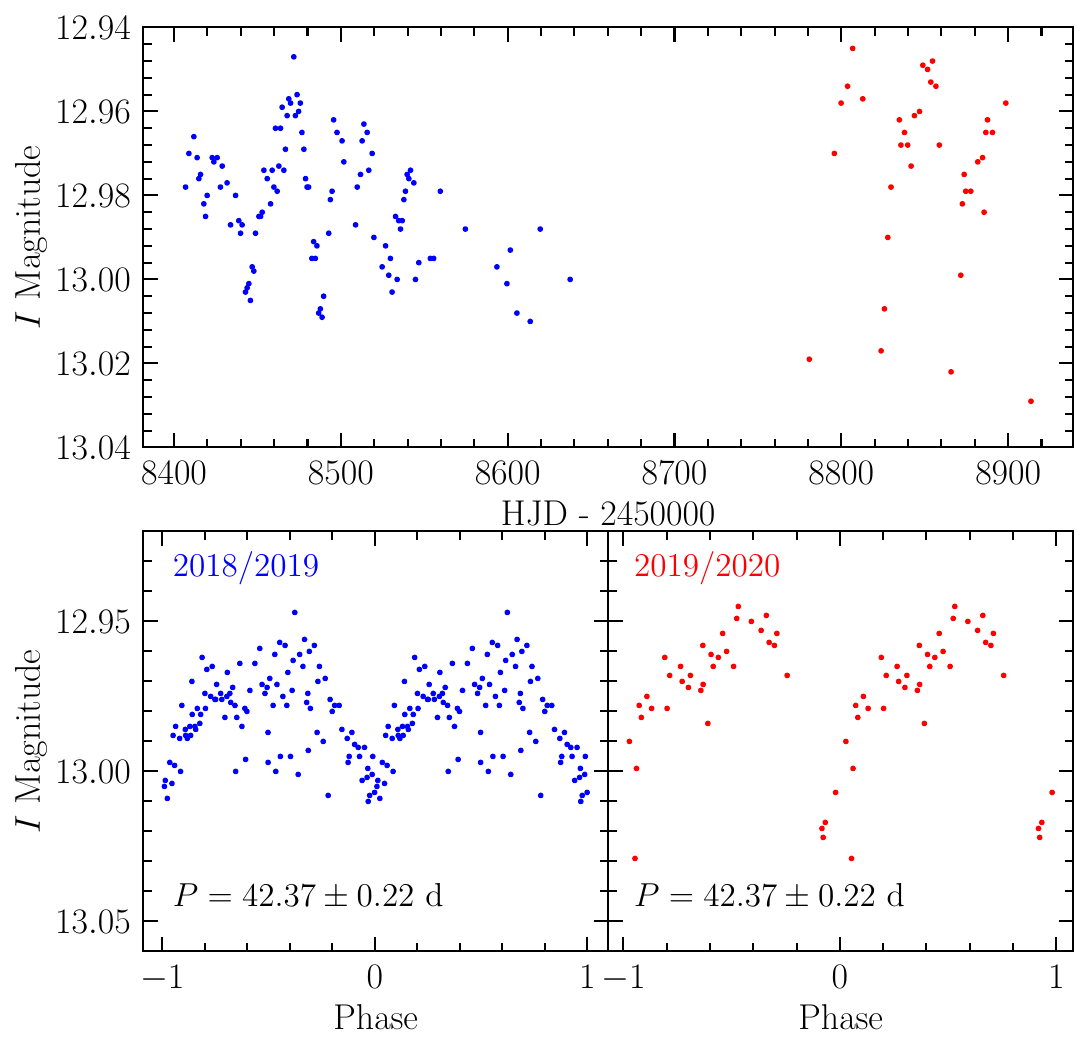}
\FigCap{Archival OGLE light curve of TIC~107150013 shows periodic variability due to starspots.}
\label{fig:lc}
\end{figure}

\section{Archival OGLE Observations of TIC~107150013} \label{sec:data}

TIC~107150013 was observed by the Optical Gravitational Lensing Experiment (OGLE; \citealt{udalski2015}) as part of its Galaxy Variability Survey between October 2018 and March 2020. The OGLE survey uses the 1.3-m telescope located at Las Campanas Observatory, Chile. The analyzed star has an internal OGLE identifier GD2353.13.32. It was observed 145 times in the $I$ band, with 25\,s exposure time. The archival light curve of TIC~107150013 is presented in the upper panel of Figure~\ref{fig:lc}.

The star shows periodic variability in the archival OGLE data with a period of $P=42.37 \pm 0.22$ days (as measured using the algorithm presented by \citealt{tatry1996}). The phase-folded light curves, separately for observing seasons 2018/2019 and 2019/2020, are shown in the lower panels of Figure~\ref{fig:lc}. The observed variability may be explained by starspots \citep{iwanek2019}. The observed period remains constant throughout OGLE light curve (and is the star's rotation period). However, the shape and amplitude of light variations change over time as the spot pattern changes. Both the rotation period and the amplitude of brightness variations in TIC~107150013 are typical for spotted stars \citep{iwanek2019}.

Even though the rotation period of the star is longer than the duration of \textit{TESS} observations, a low-amplitude, long-timescale variability can be noticed in the pre-detrended \textit{TESS} data, which are presented in Figure 2 of \citet{kunimoto2024}. {\color{black} Moreover, variability with a period of 43--44 days can be clearly detected in the publicly available light curve of TIC~107150013 from the All-Sky Automated Survey for Supernovae \citep{shappee2014,hart2023}.}

\begin{figure}[htb]
\centering
\includegraphics[width=.9\textwidth]{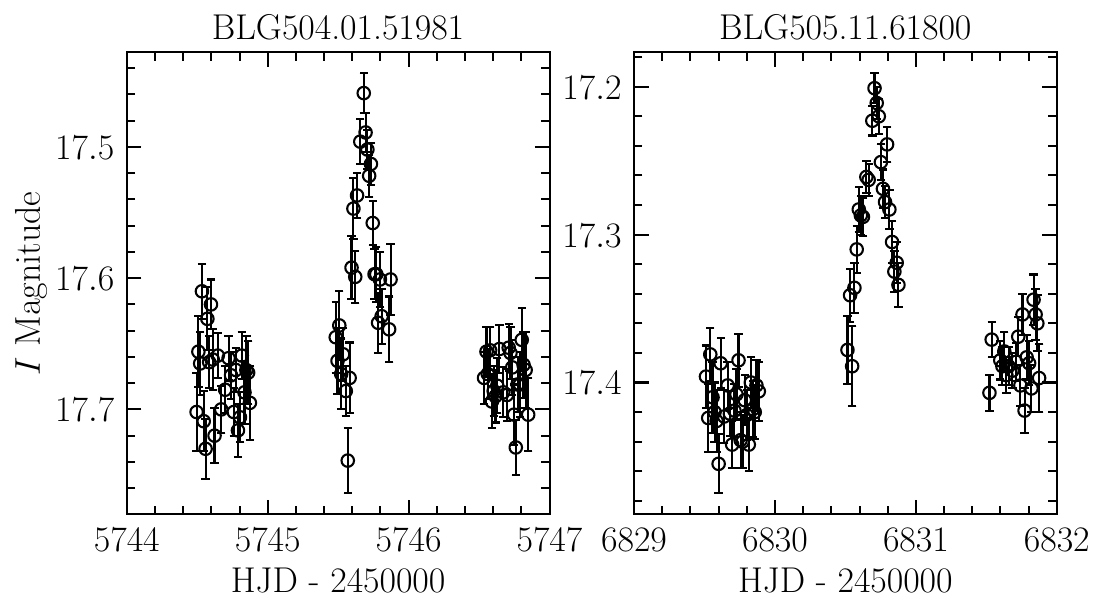}
\FigCap{Light curves of two flaring stars with symmetric flares from \citet{iwanek2019}.}
\label{fig:flares}
\end{figure}

\section{Discussion}

The presence of spots on the star's surface indicates that it is magnetically active. The stellar activity may also manifest in the form of flares \citep[e.g.,][]{iwanek2019}, which in cool stars are related to the sudden release of energy accumulated in the magnetic field. Whereas light curves of stellar flares are usually asymmetric (rapid brightening followed by a slower decline), flares may  sometimes be deceivingly symmetric.

We analyzed the light curves of 79 flaring stars published by \citet{iwanek2019} and found two objects with flares similar to that seen in TIC~107150013 (Figure~\ref{fig:flares}). Two such flaring episodes occurred for the left panel object over the last 14 years of the OGLE monitoring and just one for the right panel star. {\color{black} Taking into account diurnal and seasonal gaps in the data, the frequency of flares in these stars is $1\div 2$\,yr$^{-1}$.}
In both cases, an additional periodic variability due to starspots is present in the data. Moreover, a close inspection of light curve of TIC~107150013 seen in Figure~5 of \citet{kunimoto2024} (especially improved reductions in the left panel) reveals a fine structure of the observed brightening, similar to flickering in stellar eruptive episodes, and unexpected in smooth light curves of microlensing events. Therefore, a natural explanation of the brightening observed in the light curve of TIC~107150013 is a stellar flare, not a short-timescale, low-amplitude microlensing event. 

TIC~107150013 is a giant, and stellar flares are known to be relatively rare in such stars. In a study based on \textit{Kepler} observations, \citet{vd2017} found that about 3.18\% of giants are flaring, an incidence rate similar to that observed in main-sequence stars. However, later studies by \citet{yang2019} and \citet{olah2021} revealed that in many cases, giant oscillations or pulsations were mistaken with flares by an automated algorithm of \citet{vd2017}. These refined analyses found that only $0.3\%$ giants observed by \textit{Kepler} exhibit flares. 

Moreover, \citet{olah2021} found that the flaring activity of giants is always connected to their rotational variability. Flares were observed in giants with the photometric index $S_{\rm ph}$ higher than 0.2\%, whereas the typical inactive red giants have $S_{\rm ph} < 0.1\%$ \citep{gaulme2020}. (The photometric index is essentially the mean standard deviation of the time series data.) In the case of TIC~107150013, we measured $S_{\rm ph}=1.7\%$, which is typical for other known flaring giants \citep{olah2021}.

{\color{black} The sheer number of giants observed by \textit{TESS} makes the stellar flare interpretation much more likely than microlensing. We queried the revised \textit{TESS} Input Catalog (TIC, v8.2; \citealt{stassun2018,stassun2019,paegert2021}) to get radii and effective temperatures of $10^5$ randomly selected stars from \textit{TESS} Sector 61. We found that about 38\% are giants (using the same cut on effective temperature and radius as that adopted in Figure 11 of \citealt{stassun2019}). Therefore, the expected number of flares on giant stars is on the order of $0.3\% \times 0.38 \times 1.3 \times 10^6 \times 1 \times 27.4 / 365.25 \approx 100$, a factor of $10^5$ larger than the expected number of free-floating planets in the same data set. Most of these flaring objects were rejected by the algorithm of \citet{kunimoto2024}, except only one candidate.}

Flaring stars similar to TIC~107150013 may be a source of false positive detections in the planned space-based microlensing surveys (\textit{Roman}, \textit{ET}). The easiest method to distinguish them from genuine microlensing events would be to check the archival data for flares and periodic variability due to starspots. The contamination from flaring stars may be largest early in the missions, when the light curve archives may not be available. Another way of distinguishing flares from short-timescale microlensing events would be multiband observations: flares are known to have larger amplitudes in the bluer passbands \citep[e.g.,][]{davenport2012}, while microlensing is achromatic.

\section*{Acknowledgements}

This work has made use of data from the OGLE survey. We would like to thank OGLE observers for their contribution to the collection of the photometric data used in this paper. We thank Andrzej Udalski for his comments on the manuscript. 

%\bibliographystyle{acta}
%\bibliography{pap}

\begin{thebibliography}{23}
\providecommand{\natexlab}[1]{#1}

\bibitem[{{Davenport} \textit{et~al.}(2012)}]{davenport2012}
{Davenport}, J.R.A., \textit{et~al.} 2012, \textit{\apj}, \textbf{748}, 58.

\bibitem[{{Gaulme} \textit{et~al.}(2020)}]{gaulme2020}
{Gaulme}, P., \textit{et~al.} 2020, \textit{\aap}, \textbf{639}, A63.

\bibitem[{{Ge} \textit{et~al.}(2022)}]{ge2022}
{Ge}, J., \textit{et~al.} 2022, \textit{arXiv e-prints}, arXiv:2206.06693.

\bibitem[{{Gould} \textit{et~al.}(2022)}]{gould2022}
{Gould}, A., \textit{et~al.} 2022, \textit{\jkas}, \textbf{55}, 173.

\bibitem[{{Hart} \textit{et~al.}(2023)}]{hart2023}
{Hart}, K., \textit{et~al.} 2023, \textit{arXiv e-prints}, arXiv:2304.03791.

\bibitem[{{Iwanek} \textit{et~al.}(2019)}]{iwanek2019}
{Iwanek}, P., \textit{et~al.} 2019, \textit{\apj}, \textbf{879}, 114.

\bibitem[{{Johnson} \textit{et~al.}(2020)}]{johnson2020}
{Johnson}, S.A., \textit{et~al.} 2020, \textit{\aj}, \textbf{160}, 123.

\bibitem[{{Kunimoto} \textit{et~al.}(2024){Kunimoto}, {DeRocco}, {Smyth}, and
  {Bryson}}]{kunimoto2024}
{Kunimoto}, M., {DeRocco}, W., {Smyth}, N., and {Bryson}, S. 2024,
  \textit{arXiv e-prints}, arXiv:2404.11666.

\bibitem[{{Mr{\'o}z} and {Poleski}(2023)}]{mroz_poleski_2024}
{Mr{\'o}z}, P., and {Poleski}, R. 2023, \textit{arXiv e-prints},
  arXiv:2310.07502.

\bibitem[{{Mr{\'o}z} \textit{et~al.}(2017)}]{mroz2017}
{Mr{\'o}z}, P., \textit{et~al.} 2017, \textit{\nat}, \textbf{548}, 183.

\bibitem[{{Mr{\'o}z} \textit{et~al.}(2020)}]{mroz2020c}
{Mr{\'o}z}, P., \textit{et~al.} 2020, \textit{\apjs}, \textbf{249}, 16.

\bibitem[{{Ol{\'a}h} \textit{et~al.}(2021)}]{olah2021}
{Ol{\'a}h}, K., \textit{et~al.} 2021, \textit{\aap}, \textbf{647}, A62.

\bibitem[{{Paegert} \textit{et~al.}(2021)}]{paegert2021}
{Paegert}, M., \textit{et~al.} 2021, \textit{arXiv e-prints}, arXiv:2108.04778.

\bibitem[{{Schwarzenberg-Czerny}(1996)}]{tatry1996}
{Schwarzenberg-Czerny}, A. 1996, \textit{\apjl}, \textbf{460}, L107.

\bibitem[{{Shappee} \textit{et~al.}(2014)}]{shappee2014}
{Shappee}, B.J., \textit{et~al.} 2014, \textit{\apj}, \textbf{788}, 48.

\bibitem[{{Stassun} \textit{et~al.}(2018)}]{stassun2018}
{Stassun}, K.G., \textit{et~al.} 2018, \textit{\aj}, \textbf{156}, 102.

\bibitem[{{Stassun} \textit{et~al.}(2019)}]{stassun2019}
{Stassun}, K.G., \textit{et~al.} 2019, \textit{\aj}, \textbf{158}, 138.

\bibitem[{{Sumi} \textit{et~al.}(2023)}]{sumi2023}
{Sumi}, T., \textit{et~al.} 2023, \textit{\aj}, \textbf{166}, 108.

\bibitem[{{Udalski} \textit{et~al.}(2015){Udalski}, {Szyma{\'n}ski}, and
  {Szyma{\'n}ski}}]{udalski2015}
{Udalski}, A., {Szyma{\'n}ski}, M.K., and {Szyma{\'n}ski}, G. 2015,
  \textit{\actaa}, \textbf{65}, 1.

\bibitem[{{Van Doorsselaere} \textit{et~al.}(2017){Van Doorsselaere},
  {Shariati}, and {Debosscher}}]{vd2017}
{Van Doorsselaere}, T., {Shariati}, H., and {Debosscher}, J. 2017,
  \textit{\apjs}, \textbf{232}, 26.

\bibitem[{{Yang} and {Liu}(2019)}]{yang2019}
{Yang}, H., and {Liu}, J. 2019, \textit{\apjs}, \textbf{241}, 29.

\bibitem[{{Yang} \textit{et~al.}(2024){Yang}, {Zang}, {Gan}, {Kuang}, {Gould},
  and {Mao}}]{yang2024}
{Yang}, H., {Zang}, W., {Gan}, T., {Kuang}, R., {Gould}, A., and {Mao}, S.
  2024, \textit{arXiv e-prints}, arXiv:2405.02279.

\bibitem[{{Zhu} and {Dong}(2021)}]{zhu_dong_2021}
{Zhu}, W., and {Dong}, S. 2021, \textit{\araa}, \textbf{59}, 291.

\end{thebibliography}

\end{document}